\documentclass[doublecol]{epl2} 
\usepackage{amsfonts}
\usepackage{amsmath}
\usepackage{hyperref}
\hypersetup{colorlinks=true,linkcolor=blue,citecolor=blue,filecolor=blue,urlcolor=blue,breaklinks=true}
\usepackage{subcaption}
\title{Effect of a critical magnetic field on the control of scalar neutral boson pair production in the context of Lorentz-symmetry violation}
\shorttitle{Effect of a critical magnetic field on the control of scalar neutral boson pair production} 

\author{Andrés G. Jirón.\inst{1}}
\author{Andrés G. Jirón\inst{1} \and Angel E. Obispo \inst{1,2} \and J. Daniel Espinoza Loayza\inst{3}, Juan Carlos Quispe\inst{1} \and L. B. Castro\inst{4}.}
\shortauthor{Andrés G. Jirón\etal}

\institute{                    
  \inst{1} Universidad Tecnológica del Perú (UTP), Lima, Perú}

\institute{                    
  \inst{1} Universidad Tecnológica del Perú (UTP), Lima, Perú\\
  \inst{2} Universidad Privada del Norte (UPN), Lima, Perú\\
  \inst{3} Universidad Nacional del Callao (UNAC), Bellavista, Callao, Perú\\
  \inst{4} Universidade Federal do Maranh\~ao (UFMA), S\~ao Luís, MA, Brazil}

\abstract{This study investigates the production of neutral scalar boson pairs in static electromagnetic fields resulting from Lorentz-symmetry violation (LSV), with a focus on the parity-even sector of the CPT-even photon sector in the Standard Model Extension (SME). Utilizing a cross-configuration involving inhomogeneous static electric fields and homogeneous static magnetic fields, the analysis of the probability of bosons pair production identifies three different regimes determined by critical magnetic field. Below the critical value, creation is exponentially suppressed; at the critical value, the number density of created bosons remains constant, and above the critical field, there is exponential amplification. This  behavior prompts an additional investigation using von Neumann entanglement entropy to analyze fluctuations in the bosonic vacuum.}
\begin{document}
\maketitle
\section{INTRODUCTION}
The Sauter-Schwinger effect \cite{sauter1931verhalten,schwinger1951gauge,heisenberg1936folgerungen} occurs when an intense electric field induces spontaneous production of particle-antiparticle pairs, such as electrons and positrons, from vacuum energy. This phenomenon manifests itself in situations where the intensity of a static homogeneous electric field exceeds its critical value $E_{sch} \approx 1.30 \times 10^{18} \, \text{V/m}$, resulting in the vacuum undergoing a reconfiguration of its properties and giving rise to the spontaneous creation of virtual particles. 

Introducing a homogeneous static magnetic field to the system has demonstrated that, while this field cannot independently generate particles, its effects decrease pair production rate \cite{lin1999electron}. When the fields are neither homogeneous nor static, the theoretical techniques for analyzing pair production become more complex. Methods such as perturbative expansions \cite{best1992multiplicity}, $S$-matrix theory \cite{dyson1949s}, path integrals \cite{affleck1982pair} and numerical simulations in space-time \cite{kohlfurst2022sauter}, are challenged by the increasing demands of the system. However, for specific field configurations, the production of fermion pairs has been successfully analyzed \cite{kim2007improved,jiang2013enhancement,su2012suppression,tanji2009dynamical,su2012magnetic}. For example, in \cite{su2012suppression,su2012magnetic}, a configuration formed by a magnetic and electric field perpendicular to each other was studied. 
In this case, the magnetic field induces a cyclotron motion in the newly produced pairs, driving an electron back to the creation region. However, due to the action of the Pauli blocking principle, which prevents a newly generated electron from occupying the same state, the fermion pair production is significantly suppressed. As is evident, this phenomenon is typical of particles that obey the Fermi-Dirac statistics. Thus, in a hypothetical scenario, if bosons replaced fermions, it is believed that the boson pair production would be amplified rather than suppressed \cite{lv2013pair}. Regarding the process of boson pair production, it is essential to note that the mass of the lightest boson, the $\pi$ meson, is 270 times greater than that of the electron. Consequently, since the critical intensity of the fields depends on the square of the rest mass of the created particle, it is estimated that the intensity of the fields should be about 105 times greater than the Schwinger limit for fermions \cite{wagner2010exponential}. However, although this energy regime is currently inaccessible to the most powerful lasers, the process of boson production continues to be the subject of many interesting theoretical investigations \cite{lv2013pair,wagner2010bosonic,wagner2010exponential}. 

In this paper, the production of pairs of neutral scalar bosons in static electromagnetic fields is studied. These fields are induced by Lorentz-symmetry violation (LSV).  Specifically, the parity-even sector of the CPT-even photon sector of the Standard Model Extension (SME) \cite{casana2011dimensional,casana2010parity} will be chosen to define a cross configuration of an inhomogeneous static electric field and a homogeneous static magnetic field. In this context, it will be shown that the process of boson pair production is divided into three distinct regimes, which are determined by the presence of a critical magnetic field. Thus, the production process is exponentially suppressed if the magnetic field is less than its critical value. If the magnetic field equals its critical value, the number of produced bosons remains constant. Finally, the number of produced bosons is exponentially amplified for values greater than the critical field. This behavior is quite unexpected and exciting, so we will also study the von Neumann entanglement entropy to analyze the behavior of the fluctuations of the bosonic vacuum.

The remainder of this paper is organized as follows. In the next section, we solve the Klein-Gordon equation influenced by a crossed arrangement of a radial electric field and an axial magnetic field, represented through a non-minimal coupling based on the Lorentz-symmetry violation. In the third section, we use the Bogoliubov transformations to obtain analytical expressions associated with the probability of boson production. In the fourth section, we analyze the von Neumann entanglement entropy and the bosonic vacuum fluctuations under the influence of the critical magnetic field. Finally, we conclude in the fifth section.

\section{Klein–Gordon equation  in the background of Lorentz-symmetry violation}
Starting from the CPT-even sector of the SME \cite{casana2011dimensional,casana2010parity}, the relativistic quantum dynamics of a massive neutral scalar particle is examined under the effects of LSV. This will be achieved by introducing the following non-minimal coupling  
\begin{equation}
\hat{p}_\mu \hat{p}^\mu \rightarrow \hat{p}_\mu \hat{p}^\mu -\frac{g}{4}(K_F)_{\mu \nu \alpha \beta} F^{\mu \nu} F^{\alpha \beta},
\end{equation}
where $g$ is a coupling constant, $F^{\mu \nu}$ stands for the electromagnetic tensor, and $(K_F)_{\mu \nu \alpha \beta}$ is the tensor governing Lorentz violation in CPT-even electrodynamics within the SME framework. This tensor, which has $19$ components, exhibits the same symmetries as the Riemann tensor. These symmetries can be expressed in terms of $4$ matrices of size $3 \times 3$, defined as 
\begin{equation}
    (\kappa_{DE})_{jk}=-2(K_F)_{0j0k},
\end{equation} 
\begin{equation}
    (\kappa_{HB})_{jk}=\frac{1}{2}(K_F)^{pqlm}\varepsilon_{jpq} \varepsilon_{klm} , 
\end{equation}
\begin{equation}
(\kappa_{DB})_{jk}=-(\kappa_{HE})_{kj}= (K_F)^{0jpq}\varepsilon_{kpq}.
\end{equation}
Here, the symmetric $\kappa_{DE}$ and $\kappa_{HB}$, with 11 independent components, comprise the even-parity sector. In contrast, the $\kappa_{DB}$ and $\kappa_{HE}$ matrices, with eight components and no inherent symmetry, constitute the odd-parity sector of the $(K_F)_{\mu \nu \alpha \beta}$ tensor. Consequently, the dynamics of neutral scalar particles with mass $m$  under Lorentz violation effects are governed by the following Klein-Gordon equation in natural units
\begin{equation}
\begin{split}
&\hat{p}_\mu \hat{p}^\mu \psi +\frac{g}{2}(\kappa_{DE})_{jp} E^{j} E^{p} \psi+ g(\kappa_{DB})_{j\omega} E_{j}  B^\omega \psi\\& -\frac{g}{2}(\kappa_{HB})_{bc} B^bB^c \psi = m^2 \psi ,
\end{split}  \label{eq2}
\end{equation}
where $E_i= F_{0i}$ and $B_i=\frac{1}{2}\epsilon_{ijk}F^{jk}$ represent the electric and magnetic fields, respectively.  The following configuration will be adopted for the non-null components of the tensor $(K_F)_{\mu \nu \alpha \beta}$
\begin{equation}
    g(\kappa_{DE})_{rr}=-\kappa_1 , \quad g(\kappa_{DB})_{rz}=\kappa_3,
    \label{constant}
\end{equation}
where $\kappa_1$ and $\kappa_3$ are positive constants. 
In that scenario, the background of Minkowski space-time is considered in cylindrical coordinates
\begin{equation}
ds^2=-dt^2+dr^2+r^2 d\varphi^2+dz^2.
\end{equation}
In this context, based on the study of induced electric dipole moment systems \cite{bakke2014persistent,furtado2006landau} and other scenarios that involve LSV  \cite{vitoria2021effects,vitoria2020massive}, it is possible to define the following configuration for the crossed electric and magnetic fields
\begin{equation}
\Vec{E}= \frac{\lambda}{r} \, \hat{r} , \quad \Vec{B}=B \, \hat{z} , \label{lsv-elec-mag}
\end{equation}
where $\lambda$ is a linear electric charge density, $B$ is a constant magnetic field, and $\hat{r}$ and $\hat{z}$ are unit vectors in the radial direction and $z$ direction, respectively. Thus, considering the expressions given in (\ref{constant}) and (\ref{lsv-elec-mag}), the equation (\ref{eq2})  can be rewrite as
\begin{equation}
 \Big[- \frac{\partial^2}{\partial t^2}+  \nabla^2  +\frac{\kappa_1}{2} \frac{\lambda^2}{r^2}-\kappa_3 \frac{\lambda B}{r}-m^2\Big]\psi =0
 \label{eq3} .
\end{equation}
Here, $\nabla^2$ is the Laplacian operator in the cylindrical coordinate system. Given the cylindrical symmetry of the equation (\ref{eq3}), it is possible to choose the following ansatz
\begin{equation}
 \psi(r)= e^{-i\omega t} e^{i l \varphi} e^{i k_z z} \, \frac{\phi(r)}{\sqrt{r}}  , \label{ansatz}
\end{equation}
where $\omega$ represents the energy of the system, $l$ is the eigenvalue of the angular momentum operator, and $k_z$ is the wave-number in the $z$-direction. In this work, we consider the case where the dynamics of the boson are confined to the plane ($k_z=0$), obtaining 
\begin{equation}
    \frac{d^2\phi(r)}{dr^2}+\left[\tilde{\omega}^2 - \frac{\delta}{r}-\frac{\gamma^2_l-\frac{1}{4} }{r^2} \right]\phi(r) =0 ,\label{eq4}
\end{equation}
with
\begin{equation}
\tilde{\omega}^2=\omega^2-m^2, \qquad \delta=  \kappa_3 \lambda B , \label{delta}
\end{equation}
\begin{equation}  
    \gamma_l=  \sqrt{l^2- \frac{\kappa_1 \lambda^2}{2}}. \label{false-gamma}
\end{equation}
When $\delta > 0$ and $\frac{\kappa_1 \lambda^2}{2} \leq l^2$, equation (\ref{eq4}) represents the Schrödinger equation for the repulsive Coulomb-like potential. However, if $\frac{\kappa_1 \lambda^2}{2} \geq l^2$, the potential structure in (\ref{eq4}) is transformed into that of a well and $\gamma_l$ becomes imaginary, akin to the problem of spontaneous pair production of particles induced by a Coulomb field \cite{reinhardt1977quantum,khalilov2009spontaneous,khalilov1998fermion}. In the interest of analyzing the production of scalar bosons, a redefinition of $\gamma_l$ is performed
\begin{equation}
    \gamma_l=i\bar{\gamma}_l= i \sqrt{\frac{\kappa_1}{2}} \sqrt{\lambda^2-\lambda^{2}_{l}
      }  , \label{gammal}
\end{equation}
where
\begin{equation}
\lambda_{l}=  \sqrt{\frac{2}{\kappa_1}} l, \label{minv}
\end{equation}
under the condition $|\lambda|\geq|\lambda_{l}|$. Given the aforementioned modifications and with $z = -2 i \tilde{\omega} r$, equation (\ref{eq4}) is transformed as follows
\begin{equation}
    \frac{d^2 \phi}{dz^2} + \left[-\frac{1}{4}- \frac{i\eta}{z}  +\frac{\bar{\gamma}^2_l+1/4}{z^2} \right]\phi =0 \label{wittaker}
\end{equation}
where $\eta$ is define as
\begin{equation}
\eta=\frac{\delta}{2\tilde{\omega}},
\end{equation}
with $\tilde{\omega}\in \mathbb{R}$. The second-order differential equation given in (\ref{wittaker}) is the so-called Whittaker equation, which admits two linearly independent regular solutions, given by 
\begin{equation}
\phi= C_1 M_{-i\eta,i\bar{\gamma}_l}(z) + C_2 W_{-i\eta,i\bar{\gamma}_l}(z) , \label{sol-1}
\end{equation}
where the first solution is bounded at $z=0$, while the second is bounded at $|z| \rightarrow \infty$. The next section will use these solutions to analyze the pair production process.

\section{Bogoliubov transformation and probability of scalar bosons pair production in the Lorentz-violating background} 
\begin{figure}[tb!]
\centering
\begin{minipage}{0.45\textwidth}
  \centering
  \includegraphics[scale=0.75]{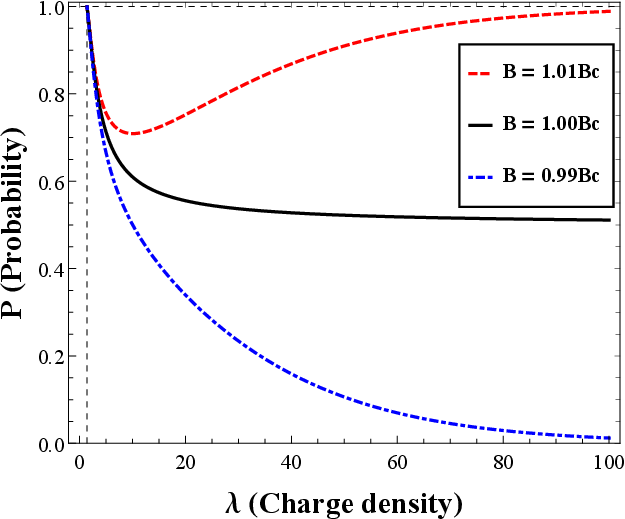}
  \caption{Probability of bosons pair production as a function of charge density $\lambda$ for different magnetic field values $B$, where $l=1$, $\tilde{\omega}=1{.}12$ and $\kappa_1=\kappa_3=1$.}
  \label{Fig:prob-Elec}
\end{minipage}
\end{figure}

In accordance with the prescription used in \cite{padmanabhan1991quantum,haouat2014influence,hamil2020pair,belbaki2023influence}, the in-out states, as described in equation (\ref{sol-1}), are defined as follows:
\begin{equation}
\phi^+_{in}(r)=C_1\,M_{-i\eta,i\bar{\gamma}_l}(-2i\tilde{\omega} r ),
\label{sol-11}
\end{equation}
\begin{equation}
\phi^+_{out}(r)=C_2\, W_{-i\eta,i\bar{\gamma}_l}(-2i\tilde{\omega} r ),
\end{equation}
\begin{equation}
\phi^-_{in}=\left[\phi^+_{in}(r) \right]^*=C^*_1\,M_{i\eta,-i\bar{\gamma}_l}(2i\tilde{\omega} r ),
\end{equation}
\begin{equation}
\phi^-_{out}=\left[\phi^+_{out}(r)\right]^*=C^*_2\, W_{i\eta,-i\bar{\gamma}_l}(2i\tilde{\omega} r ). \label{sol-44}
\end{equation}
The indices $\pm$ denote the modes with positive and negative frequency, respectively. These states are connected through the so-called Bogoliubov transformations,  which are obtained using the following relations of the Whittaker functions \cite{lozier2003nist} 
\begin{equation}
M_{k,\mu}(ze^{\pm i \pi}) = e^{\pm i \pi (1/2+\mu)}  M_{-k,\mu}(z), 
\end{equation}
\begin{equation}
\begin{split}
W_{k,\mu}(z) &= \frac{\Gamma(-2\mu)}{\Gamma(1/2-\mu-k)}M_{k,\mu}(z)\\&+ \frac{\Gamma(2\mu)}{\Gamma(1/2+\mu-k)}M_{k,-\mu}(z) .
\end{split}
\end{equation}
\begin{equation}
\begin{split}
\frac{M_{k,\mu}(z)}{\Gamma(1+2\mu)}  &= \frac{e^{\pm(\kappa-\mu-1/2)i\pi}}{\Gamma(1/2+\mu+k)}W_{k,\mu}(z)\\&+ \frac{e^{\pm \kappa i\pi}}{\Gamma(1/2+\mu-k)}W_{-k,\mu}(e^{\pm i \pi}z) .
\end{split}
\end{equation}
The relation between the states $\phi^{\pm}_{in}(r)$ and $\phi^{\pm}_{out}(r)$ are given by
\begin{equation}
\phi^{+}_{in}(r)=\alpha^* \phi^+_{out}(r)-\beta\, \phi^-_{out}(r).
\end{equation}
\begin{equation}
\phi^+_{out}(r)= \alpha\, \phi^+_{in}(r)+\beta\, \phi^-_{in}(r),
\end{equation}
where the Bogoliubov coefficients $\alpha$ and $\beta$ are expressed as
\begin{equation}
\alpha= \frac{C_2}{C_1}\,  \frac{\Gamma(-2i\bar{\gamma}_l)}{\Gamma(1/2-i\bar{\gamma}_l +i\eta)},  \label{coefAB-2}
\end{equation}
\begin{equation}
\beta=\frac{C^*_2}{C_1} \, \frac{\Gamma(2i\bar{\gamma}_l)}{\Gamma(1/2-i\bar{\gamma}_l-i\eta)} e^{-\pi\bar{\gamma}_l}e^{i\pi/2},  \label{coefAB-1}
\end{equation}
which satisfies the conditions
\begin{equation}
|\alpha|^2-|\beta|^2=1  ,\label{rela}
\end{equation}
with 
\begin{equation}
\frac{|C_2|^2}{|C_1|^2}=2\,\bar{\gamma}_l \, e^{\pi(\eta+\bar{\gamma}_l)}.
\end{equation}
A straightforward calculation allows the establishment of a connection between the states $\phi^\pm_{in}(r)$ and $\phi^\pm_{out}(r)$ and the creation/annihilation operators in quantum field theory
\begin{equation}
a_{in}(r)=\alpha^* \,b_{out}(r) - \beta  \, b^\dagger_{out} (r),
\end{equation}
\begin{equation}
b_{out}(r)=\alpha \,a_{in} (r) + \beta \, a^\dagger_{in} (r).
\end{equation} 
Consequently, using the coefficients $\alpha$ and $\beta$, along with the relations involving gamma functions, it is possible to obtain an expression for the probability of scalar bosons pair production  in the following way
\begin{equation}
P= \frac{|\beta|^2}{|\alpha|^2} = \frac{\cosh[\pi(\bar{\gamma}_l+\eta)]}{\cosh[\pi(\bar{\gamma}_l-\eta)]} e^{-2\pi \bar{\gamma}_l}. \label{bogo-p} 
\end{equation}
Note that the equation (\ref{bogo-p}) exhibits a decreasing behavior due to the presence of the negative exponential, which depends on both the intensity of the charge density $\lambda$ and the value of $l$. On the other hand, it reaches a maximum probability of $P=1$ when $\lambda=\lambda_l$. It is interesting to highlight that this behavior persists even in the absence of a magnetic field ($B=0$), thereby re-expressing equation (\ref{bogo-p}) as follows
\begin{equation}
P_0= \exp\left[-2\pi \sqrt{\frac{\kappa_1}{2}} \left(\sqrt{\lambda^2-\lambda_l^2}\right)\right], \label{probabilityc}
\end{equation}
where  it can be seen that $\lambda_l$ genuinely represents a threshold value, marking the point at which the boson pair production process begins. Nevertheless, the presence of the magnetic field also plays a significant role in this process, even when the electric field intensity far exceeds its threshold value ($\lambda \gg \lambda_{l}$). In this scenario, the expression for the probability (\ref{bogo-p}) is approximated as follows 
 \begin{equation}
 P \approx \frac{1+e^{- \zeta \lambda(B+B_c)}}{1+e^{- \zeta \lambda(B-B_c)}} ,\label{probabilityb}
\end{equation}
where $\zeta= \frac{\pi \kappa_3}{\tilde{\omega}}$ and $B_c$ is given by  
\begin{equation}
B_{c}= \frac{\sqrt{2\kappa_1}}{ \kappa_3 } \, \tilde{\omega}. \label{bcrit}
\end{equation} 
From (\ref{probabilityb}), three asymptotic approximations for the probability can be identified based on the value of $B_c$, as depicted in Figure \ref{Fig:prob-Elec}. When $B<B_c$, the probability consistently decreases, converging to zero. At critical point $B = B_c$, the probability converges to $\frac{1}{2}$. Conversely, for $B > B_c$, the probability increases approaching one. Therefore, when $\lambda \gg \lambda_{l}$, the magnetic field acts as a transition parameter between two distinct states: production and vacuum, where $B_c$ plays the role of critical parameter. 
In the case of other values of $\lambda$, particularly at $B=0.99 \, B_c$ (dotted blue line), $B=1.00 \, B_c$ (solid black line), and $B=1.01 \, B_c$ (dot-dashed red line), a notable convergence among the three probability curves occurs precisely at the critical threshold $\lambda_l=\sqrt{2}$ (vertical dashed line), which is determined by equation (\ref{minv}). At $B=1.00 \, B_c$, the probability decays similarly to the case of $B=0.99 \, B_c$, yet it converges to a constant value of $P=1/2$, suggesting equiprobability for all possible quantum states. Conversely, for $B=1.01 \, B_c$, the probability initially decreases to a minimum point at $\lambda_{min}=10.07$ and thereafter gradually rises until it approaches one. By computing the first and second derivatives of (\ref{bogo-p}), additional minimum points can be identified. For example, at $B=1.08 \, B_c$, the minimum point is located at $\lambda_{min}=3.74$, with a probability $P=0.93$. If the magnetic field is slightly increased to $B=1.25 \, B_c$, the minimum point shifts to $\lambda_{min}=2.36$ and $P=0.99$. These results indicate that as the magnetic field gradually increases, $\lambda_{min}$ approaches the threshold value $\lambda_l$ and the probability approaches one.

\begin{figure}[tb!]
\centering
\begin{minipage}{0.45\textwidth}
  \centering
  \includegraphics[scale=0.76]{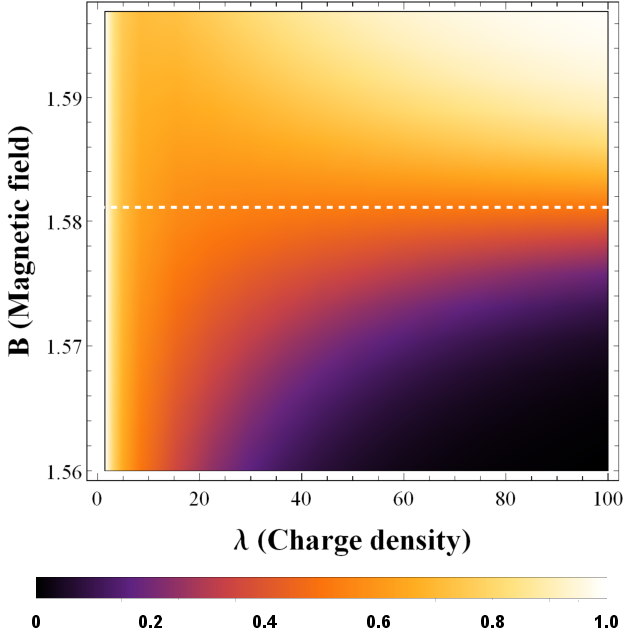}
  \caption{Density plot of the probability of bosons pair production as a function of charge density $\lambda$ and magnetic field values $B$, where $l=1$, $\tilde{\omega}=1{.}12$ and $\kappa_1=\kappa_3=1{.}0$.}
  \label{Fig:prob-Elec2}
\end{minipage}
\end{figure}

Other scenarios involving different values of magnetic field and charge density can be explored through a density plot displaying the probability of boson pair creation, as depicted in Figure \ref{Fig:prob-Elec2}. Notably, below the critical magnetic field $B_c=1.584$ (horizontal white dashed line), two distinct regions are evident, indicated by light and dark colors, which represent high and low probabilities, respectively. Here, a noticeable trend emerges: a higher probability region occurs for charge densities below $\lambda \approx 15$, whereas lower probability regions are observed for both higher charge densities surpassing this threshold and magnetic fields below the critical value $B_c$. In contrast, when the magnetic field exceeds the critical value, only regions of high probability exist for any given charge density. It is interesting to note that the probability of boson pair production is sensitive to slight changes in the magnetic field around the critical value $B_c$. This behavior is also observed in the profile of the number density of produced escalar-neutral bosons, which will be discussed in the next section.
\section{Boson pair production in the Lorentz-violating background} 
\begin{figure}[tb!]
\centering
\begin{minipage}{0.45\textwidth}
  \centering
  \includegraphics[scale=0.76]{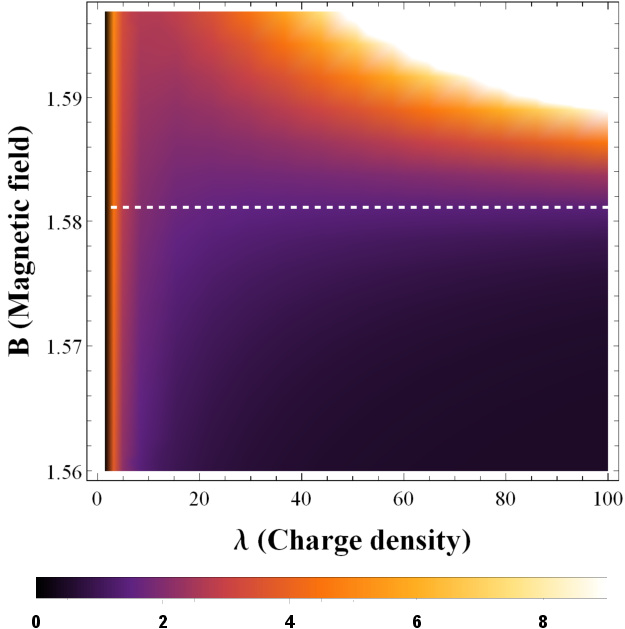}
  \caption{Density plot of the number density of created bosons as a function of charge density $\lambda$ and magnetic field values $B$, where $l=1$, $\tilde{\omega}=1{.}12$ and $\kappa_1=\kappa_3=1{.}0$.}
  \label{Fig:creacion-Elec2}
\end{minipage}
\end{figure}

To calculate the number density of created particles or number of created particles per state, the matrix element given by equation (\ref{densitty}) is employed
\begin{equation}
\hat{n}=\left<0_{\text{in}}| a^\dagger_{\text{out}} , a_{\text{out}}|0_{\text{in}} \right>=|\beta|^2. \label{densitty}
\end{equation}
Using equations (\ref{rela}) and (\ref{bogo-p}), the number density of created neutral bosons is determined as follows
\begin{equation}
\hat{n}_{l} = \frac{\cosh[\pi(\eta+\bar{\gamma}_l)]}{\sinh 2\pi \bar{\gamma}_l} e^{\pi (\eta-\bar{\gamma}_l)} .
\label{n-creation}
\end{equation}
It is noteworthy that the expression in equation (\ref{n-creation}) exhibits divergence at $\bar{\gamma}_l=0$, a condition met only when $\lambda = \lambda_l$. This divergence suggests an infinite production of bosons, which can be interpreted as an infinite number of continuous states condensed within an infinitesimal space \cite{good2019finite}. On the other hand,  when $B=0$, the number density given by Equation (\ref{n-creation}) can be represented as a Bose-Einstein distribution, expressed as
\begin{equation}
\hat{n}_{0l} = \frac{1}{e^{2\pi \sqrt{\frac{\kappa_1}{2}}\lambda}-1}.
\label{n-creation-as}
\end{equation}
It is observed that $\hat{n}_{0l}$ approaches zero as $\lambda$ significantly increases. This notable absence of created states might suggest that the energy required to generate boson-antiboson pairs is sufficiently high, resulting in minimal neutral-scalar bosons production.\\
\begin{figure}[tb!]
\centering
\includegraphics[scale=0.75]{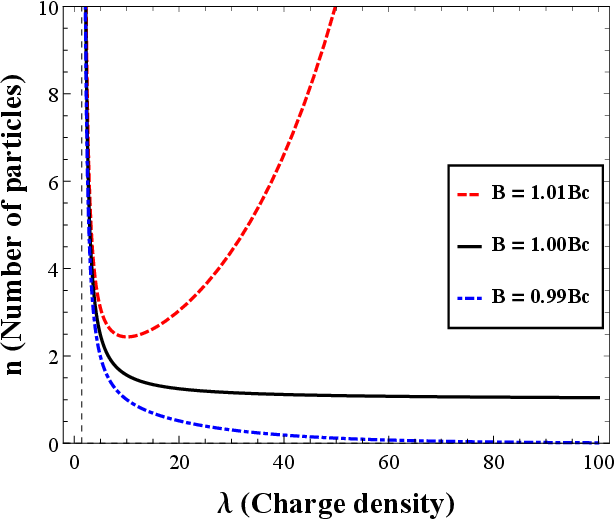}
\caption{Number density of created bosons as a function of charge density $\lambda$ for different values of the magnetic field $B$, where $l=1$, $\tilde{\omega}=1{.}12$ and $\kappa_1=\kappa_3=1{.}0$.}
\label{Fig:creacion-Elec}
\end{figure}
In Figure \ref{Fig:creacion-Elec}, the behavior of the number density of created bosons is depicted as a function of charge density, considering specific magnetic field values: $B=0.99 \, B_{c}$ (dotted blue line), $B= 1.00 \, B_{c}$ (solid black line), and $B=1.01 \, B_{c}$ (dot-dashed red line).  As observed, for $B=0.99 \, B_{c}$, the number of created bosons tends to asymptotically decay to zero for large values of $\lambda$, similar to the case where $B=0$. This exponential suppression  is unusual, as it is expected that the Pauli blocking should be responsible for amplifying the production of boson pairs in the presence of a magnetic field \cite{lv2013pair}.  For $B=1.00 \, B_{c}$, the number of bosons decreases as the charge density increases.  However, unlike the previous case, it converges to one. This behavior suggests that when the magnetic field reaches its critical value, the rate at which boson pairs are created matches the rate at which they are annihilated. For $B=1.01 B_{c}$, the number density of created bosons was suppressed below $\lambda_{min}=10.07$,  followed by a rapid increase for $\lambda > \lambda_{min}$. This phenomenon, known as Bose enhancement is a characteristic behavior observed in particle production processes governed by Bose-Einstein statistics \cite{tanji2009dynamical}. Regarding alternative scenarios encompassing various magnetic field and charge density values, Figure \ref{Fig:creacion-Elec2} presents a plot representation of the behavior of the number density of created bosons as a function of these magnitudes. Below the critical value of the magnetic field, $B_c = 1.584$ (horizontal white dashed line), there is a region represented by colors ranging from dark to even darker. This suggests that in this scenario, the number density of created bosons decreases as the charge density increases. In contrast, when the magnetic field exceeds the critical value, unlike the previous context, now a region with dark colors gradually transitioning to lighter shades is observed. This indicates a significant increase in the number density of created bosons for high values of charge density.

The results in this section illustrate the susceptibility of the pair production process to minor fluctuations around the critical magnetic field, $B_c$ induced by Lorentz-violating. This observation suggests a potential magnetic phase transition within a quantum vacuum, which alters the conventional particle production process \cite{rojas2007magnetic}. To delve deeper into this phenomenon, the subsequent section explores quantum vacuum fluctuations using the von Neumann entanglement entropy approach.

\section{von Neumann entanglement entropy in the LSV background}

\begin{figure}[tb!]
\centering
\includegraphics[scale=0.75]{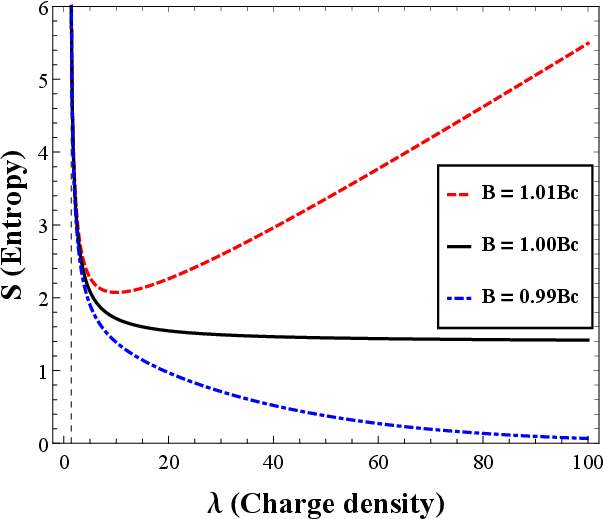}
\caption{von Neumann entropy as a function of charge density $\lambda$ for different values of the magnetic field, where $l=1$, $\tilde{\omega}=1.12$, $\kappa_1=\kappa_3=1.0$.}
\label{Fig:creacion-Elec-entropy}
\end{figure}
In few words, the von Neumann entanglement entropy quantifies the degree of entanglement within a system \cite{furuya2020generalized, lin2010quantum,ghiti2021neumann,ghiti2023quantum}. Mathematically expressed as
\begin{equation}
S=(\hat{n}+1)  \log \left(1+\hat{n}\right) - \hat{n}  \log \hat{n} \label{entropyS}.
\end{equation}
This equation characterizes the entropy based on the number density of the created boson pairs, defined in equation (\ref{n-creation}). 

\begin{figure}[tb!]
\centering
\includegraphics[scale=0.76]{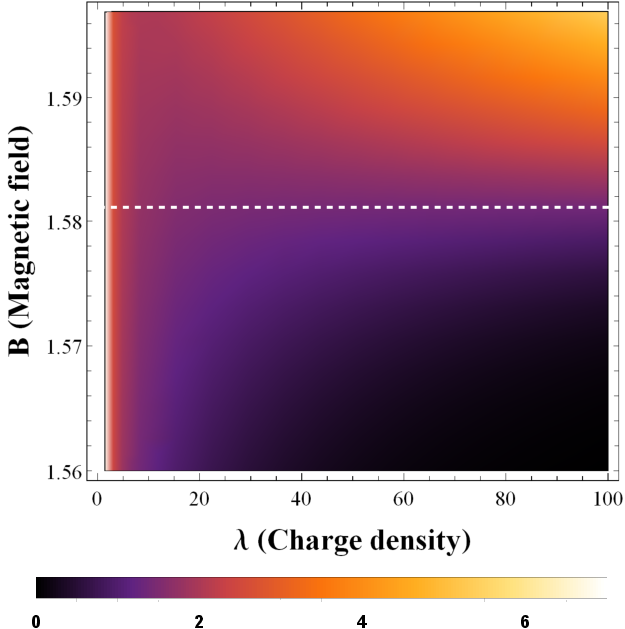}
\caption{ Density plot of the von Neumann entropy  as a function of charge density $\lambda$ and magnetic field values $B$, where $l=1$, $\tilde{\omega}=1{.}12$ and $\kappa_1=\kappa_3=1{.}0$.}
\label{Fig:creacion-Elec-entropy2}
\end{figure}

In Figure \ref{Fig:creacion-Elec-entropy}, the entropy behavior as a function of the charge density is depicted, utilizing the identical magnetic field values of $B$ that were employed in the preceding section. For $B=0.99 \, B_{c}$, the entropy asymptotically decreases to zero. This indicates that in scenarios with high charge density and magnetic fields below their critical value, quantum vacuum fluctuations become practically negligible. In such situations, the system is perfectly defined and pure \cite{anand2011shannon}.
For $B=1.00 \, B_c$, it is observed that the entropy decreases; however, it tends to stabilize at a constant value. This behavior arises because, in this scenario, the asymptotic behavior of the number of particles approaches to one, subsequently causing the entropy to converge to $S = 2 \, \log 2$. This pattern could suggest  a potential quantum equilibrium state when the critical value of $B$ is reached. For $B=1.01 B_{c}$, it is observed that the entropy decreases until it reaches a minimum located at $\lambda_{min}=10.07$, and for values above this point, the entropy increases linearly. This behavior in the entropy suggests that it approaches a state of maximum disorder and reduced predictability when the charge density increases dramatically. 
Exploring various scenarios where the magnetic field and charge density take on different values, Figure \ref{Fig:creacion-Elec-entropy2} provides a plot representation of the behavior of von Neumann entropy in relation to the aforementioned magnitudes. As observed, similar to the previous case (see Figure \ref{Fig:creacion-Elec2}), below the critical magnetic field $B_c = 1.584$ (horizontal white dashed line), there is a region ranging from a dark color to an even darker one. This suggests that in this region, there is low entanglement entropy as the charge density increases. When the magnetic field exceeds its critical value, a dark region is observed gradually lightening, indicating a slight increase in entanglement entropy as the charge density also increases.

\section{Conclusions}
This investigation focused on exploring the production of neutral scalar boson pairs within static electromagnetic fields caused by Lorentz-symmetry violation in the parity-even sector of the CPT-even photon sector in the Standard Model Extension. This was achieved by implementing a cross-configuration incorporating inhomogeneous static electric fields and homogeneous static magnetic fields. By employing the Bogoliubov transformation, an analytical expression was derived to calculate the probability of scalar boson-pair production. Although the magnetic field itself did not directly produce boson pairs, its influence not only affected the pair production process but also facilitated control over it by modulating a critical parameter, denoted as $B_c$.

By evaluating $B_c$, three distinct asymptotic approximations for the probability were identified. Below the critical threshold ($B < B_c$), the probability consistently diminishes, asymptotically approaching zero. At the critical point ($B = B_c$), the probability stabilizes at $\frac{1}{2}$. Conversely, beyond the critical threshold ($B > B_c$), the probability increases, tending toward unity. This behavior demonstrates that the magnetic field acts as a transition parameter between two distinct states: production and vacuum. Similarly, an expression for the number density of the created bosons was derived,  which revealed different behaviors. For $B<B_c$, the density asymptotically decreases. For $B=B_c$, it remained constant, while for $B>B_c$, a Bose enhancement was observed \cite{tanji2009dynamical}. On the other hand, the von Neumann entanglement entropy asymptotically observed that, for $B>B_c$, it asymptotically decreases. In the case of $B=B_c$, it tends towards a constant value, whereas for $B<B_c$, it undergoes linear growth. Finally, it is worth mentioning that if $\kappa_3<0$ is considered in the calculation of the probability, number density of the created bosons, and entanglement entropy, these would exponentially decay to zero for any value of the magnetic field $B$.

\acknowledgments
L. B. Castro acknowledges the support provided in part by funds from CNPq, Brazil, Grants No. 09126/2019-3 and 311925/2020-0, FAPEMA, and CAPES - Finance code 001. Angel E. Obispo acknowledges the financial support from the Universidad Tecnológica del Perú (UTP).

\bibliographystyle{spphys}

\begin{thebibliography}{100}
\providecommand{\url}[1]{{#1}}
\providecommand{\urlprefix}{URL }
\expandafter\ifx\csname urlstyle\endcsname\relax
  \providecommand{\doi}[1]{DOI \discretionary{}{}{}#1}\else
  \providecommand{\doi}{DOI \discretionary{}{}{}\begingroup
  \urlstyle{rm}\Url}\fi

\bibitem{sauter1931verhalten}Sauter, F., {\em Zeitschrift Für Physik}., \textbf{69}, 742-764 (1931).

\bibitem{schwinger1951gauge}Schwinger, J., {\em Phys. Rev.}., \textbf{82}, 664 (1951).

\bibitem{heisenberg1936folgerungen}Heisenberg, W. and Euler, H., {\em Zeitschrift Für Physik}., \textbf{98}, 714-732 (1936).

\bibitem{lin1999electron}Lin, Q., {\em J. Phys. G: Nucl. Part. Phys.}., \textbf{25}, 17 (1999).


\bibitem{best1992multiplicity}Best, C., Greiner, W. and Soff, G., {\em Phys. Rev. A}., \textbf{46}, 261 (1992).


\bibitem{dyson1949s}Dyson, F., {\em Phys. Rev.}, \textbf{75}, 1736 (1949).

\bibitem{affleck1982pair}Affleck, I., Alvarez, O. \& Manton, N., {\em Nucl. Phys. B}., \textbf{197}, 509-519 (1982).


\bibitem{kohlfurst2022sauter}Kohlfürst, C., Ahmadiniaz, N., Oertel, J. and Schützhold, R., {\em PRL}., \textbf{129}, 241801 (2022).


\bibitem{kim2007improved}Kim, S. and Page, D., {\em Phys. Rev. D}., \textbf{75}, 045013 (2007).

\bibitem{jiang2013enhancement}Jiang, M., Lv, Q., Sheng, Z., Grobe, R. and Su, Q., {\em Phys. Rev. A}., \textbf{87}, 042503 (2013).

\bibitem{su2012suppression}Su, W., Jiang, M., Lv, Z., Li, Y., Sheng, Z., Grobe, R. and Su, Q., {\em Phys. Rev. A}., \textbf{86}, 013422 (2012).

\bibitem{tanji2009dynamical}Tanji, N., {\em Ann. Phys.},\textbf{324}, 1691-1736 (2009).


\bibitem{su2012magnetic}Su, Q., Su, W., Lv, Q., Jiang, M., Lu, X., Sheng, Z. and Grobe, R., {\em PRL}., \textbf{109}, 253202 (2012).

\bibitem{lv2013pair}Lv, Q., Su, A., Jiang, M., Li, Y., Grobe, R. and Su, Q., {\em Phys. Rev. A}., \textbf{87}, 023416 (2013).


\bibitem{wagner2010exponential}Wagner, R., Ware, M., Su, Q. and Grobe, R.,{\em Phys. Rev. A}., \textbf{81}, 052104 (2010).


\bibitem{wagner2010bosonic}Wagner, R., Ware, M., Su, Q. and Grobe, R.,{\em Phys. Rev. A}., \textbf{81}, 024101 (2010).



\bibitem{casana2011dimensional}Casana, R., Carvalho, E. and Ferreira Jr, M., {\em Phys. Rev. D}. \textbf{84}, 045008 (2011).

\bibitem{casana2010parity}Casana, R., Ferreira Jr, M. and Silva, M., {\em Phys. Rev. D}., \textbf{81}, 105015 (2010).




\bibitem{bakke2014persistent}Bakke, K. and Furtado, C., {\em Eur. Phys. J. B}., \textbf{87} pp. 1-6 (2014).


\bibitem{furtado2006landau}Furtado, C., Nascimento, J. and Ribeiro, L.,{\em Phys. Lett. A}., \textbf{358}, 336-338 (2006).



\bibitem{vitoria2021effects}Vitória, R. and Belich, H.,{\em Eur. Phys. J. D}. \textbf{75}, 291 (2021).



\bibitem{vitoria2020massive}Vitória, R. and Belich, H.,{\em Eur. Phys. J. Plus}. \textbf{135}, 123 (2020).


\bibitem{reinhardt1977quantum}Reinhardt, J. and Greiner, W., {\em Rep. Prog. Phys.}, \textbf{40}, 219 (1977).

\bibitem{khalilov2009spontaneous}Khalilov, V., {\em Theor. Math. Phys.}, \textbf{158} (2009).

\bibitem{khalilov1998fermion}Khalilov, V.,  {\em Theor. Math. Phys.}, \textbf{116}, 956-963 (1998).


\bibitem{padmanabhan1991quantum}Padmanabhan, T. ,{\em Pramana}., \textbf{37} pp. 179-233 (1991).

\bibitem{haouat2014influence}Haouat, S. and Nouicer, K., {\em Phys. Rev. D}., \textbf{89}, 105030 (2014).

\bibitem{hamil2020pair}Hamil, B., Merad, M. and Birkandan, T., {\em Int. J. Mod. Phys. A}., \textbf{35}, 2050014 (2020).

\bibitem{belbaki2023influence} Belbaki, B. and    Bounames, A.,  {\em Int. J. Theor. Phys.} \textbf{62}, 136 (2023).

\bibitem{lozier2003nist}Lozier, D., {\em Ann. Math. Artif. Intell.}., \textbf{38} pp. 105-119 (2003).



\bibitem{good2019finite}Good, M. and Linder, E., {\em Phys. Rev. D}., \textbf{99}, 025009 (2019).


\bibitem{rojas2007magnetic}Rojas, H.,  Querts, E., {\em Int. J. Mod. Phys. D}. \textbf{16}, 165-173 (2007).

\bibitem{furuya2020generalized}Furuya, K., Lashkari, N. and Ouseph, S.,{\em JHEP}. \textbf{2020}, 1-44 (2020).



\bibitem{lin2010quantum}Lin, S., Chou, C. and Hu, B., {\em Phys. Rev. D}. \textbf{81}, 084018 (2010).

\bibitem{ghiti2021neumann}Ghiti, M., Mebarki, N. and Aissaoui, H., {\em J. Phys. Conf. Ser}. \textbf{1766}, 012023 (2021).

\bibitem{ghiti2023quantum}Ghiti, M., Aissaoui, H. and Mebarki, N., {\em Ind. J. Phys.} pp. 1-11 (2023).



\bibitem{anand2011shannon}Anand, K., Bianconi, G. and Severini, {\em Phys. Rev. E}. \textbf{83}, 036109 (2011).




\end{thebibliography}

\end{document}